\input{psfig.sty}
\documentstyle[12pt]{article}
\newfont{\feff}{cmti10}
\begin{document}

\title{Strong Turbulence in d-Dimensions }

\author{ Victor Yakhot\\
Department of Aerospace and Mechanical Engineering\\
Boston University, 
Boston, MA 02215 \\
and\\
Program in Applied and Computational Mathematics\\
Princeton University}

\maketitle

${\bf Abstract}.$
\noindent

In the limit 
$d\rightarrow\infty$ the role of pressure gradients and that of the 
incompressibility constraint  decreases, thus blurring the difference
 between transverse and longitudinal velocity correlation functions.
Using  Polyakov's expression for the dissipation anomaly
the closed equation for the probability density function is obtained. 
This model for the dissipation terms is the only one satisfying both
 equations  of motion and a set of
dynamical constraints. The resulting  equations show that when 
$d\rightarrow\infty$, the 
predictions of 
Kolmogorov theory are  exact. 
It is also shown that the $O(1/d)$ pressure effects,  producing the 
regularization of the equations of motion for the PDF are
responsible for the 
distinction between the Navier-Stokes and Burgers dynamics.

\newpage

The problem of  d-dimensional Navier-Stokes turbulence was first considered by 
Frisch, Fournier and Rose [1]  
who hoped to develop a  renormalized 
perturbation expansion with $1/d$ as a small parameter. This attempt 
failed because the $1/d$-factors appearing in the Wyld diagrammatic 
expansion due to the angular integrations  are cancelled by the $O(d)$-
multipliers,  resulting from the summation over the $d$ components of the 
velocity field. In this paper,   using the
nonperturbative approach, 
we will revisit the problem of 
$d$-dimensional turbulence

\noindent The equations of motion are (density $\rho\equiv 1$):

\begin{equation}
\partial_{t}v_{i}+v_{j}\partial_{j}v_{i}=-\partial_{i}p+\nu\nabla^{2}v_{i}+f_{i}
\end{equation}
\noindent and
\begin{equation}
\partial_{i}v_{i}=0
\end{equation}

\noindent where $\bf f$ is a forcing function mimicking the 
large-scale turbulence production mechanism and in a statistically 
steady state the mean pumping rate $P
=\overline{{\bf f\cdot v}}={\cal E}=O(1)$
where 
${\cal E}$ is the mean dissipation rate. In what follows 
we will also deal with the local dissipation rate

$${\cal E}({\bf x})=\nu(\frac{\partial v_{i}}{\partial x_{j}})^{2}$$

\noindent so that ${\cal E}=\overline{{\cal E}(x)}$.
The mean dissipation rate is independent
on the space dimensionality $d$.

Among other results, Kolmogorov theory produced a
 clear distinction between longitudinal and transverse structure
 functions: Consider two points
${\bf x}$ and ${\bf x'}$ and define ${\bf r}={\bf x-x'}$. Assuming that
the $x$-axis is paralel to the displacement vector ${\bf r}$,  one
can find that in the inertial range [2],[3]:

\begin{equation}
\frac{1}{r^{d+1}}\partial_{r}r^{d+1}S_{3}=-\frac{12}{d}{\cal E}
\end{equation}

\noindent giving  
\begin{equation}
S_{3}=\overline{(\Delta u)^{3}}
\equiv\overline{(u(x')-u(x))^{3}}\approx -\frac{12}{d(d+2)}r
\end{equation}
\noindent and
\begin{equation}
S^{t}_{3}=\overline{(\Delta v)^{3}}
\equiv\overline{(v(x')-v(x))^{3}}\approx 0
\end{equation}

\noindent where $u$ and $v$ are the components of  velocity field paralel
and perpendicular to the $x$-axis  (vector ${\bf r})$. These two relations 
show that the energy flux in the $d$-dimensional turbulence 
is dominated by the longitudinal velocity fluctuations only and that the
remaining $d-1$
transverse components do not directly participate in the energy transfer.
It is easy to show
that the probability density 
$P(\Delta u,r)=P(-\Delta u,-r)$,
while 
$P(\Delta v,r)=P(-\Delta v,r)$.
Thus, the time-reversal symmetries  of 
 transverse and longitudinal velocity  fluctuations are 
very different. This difference is lost in the Fourier-transforms 
of the relations (4) and (5),  both giving $S_{3}(k)=S_{3}^{t}(k)=0$ for $k\neq 0$.
That is why the  Wyld expansion,  dealing  with all components
of  velocity field on an equal footing,
is unable to make a crucial distinction  between longitudinal
 and transverse velocity
fluctuations.  The relations (4), (5) also show that in the limit 
$d\rightarrow\infty$ the asymmetry of the probability density 
function, responsible for non-zero values of the odd-order moments 
 decreases.

\noindent  
We consider the $N$-point generating function:
\begin{equation}
Z=<e^{\lambda_{i}\cdot {\bf v(x_{i})}}>
\end{equation}

\noindent where the vectors ${\bf x_{i}}$ define  the positions of the points 
denoted $1\leq i \leq N$. 
Using the incompressibility condition,
the equation for $Z$ can be  written:
\begin{equation}
\frac{\partial Z}{\partial t}+\frac{\partial^{2} Z
}{\partial \lambda_{i,\mu}\partial  x_{i,\mu}}=I_{f}+I_{p}+D
\end{equation}

\noindent with 
\begin{equation}
I_{f}=\sum_{j} <{\bf \lambda_{j}\cdot f(x_{j})}e^{\lambda_{i}u(x_{i})}>
\end{equation}
\begin{equation}
I_{p}=-\sum_{j}\lambda_{j}<e^{\lambda_{i}u(x_{i})}\frac{\partial p(x_{j})}{\partial x_{j}}>
\end{equation}
\noindent and  
\begin{equation}
D=\nu \sum_{j} \lambda_{j}<e^{\lambda_{i}u(x_{i})}\frac{\partial^{2} u(x_{j})}
{\partial x_{j}^{2}}>
\end{equation}

In what follows we will be mainly interested 
in the probability density function of the two-point velocity
differences which is ontained from (7)-(10), setting 
$\bf{\lambda_{1}+\lambda_{2}}=0$ (sse Refs. [4]-[6]), 
so that 
\begin{equation}
Z=<exp{(\bf{\lambda\cdot U})}>
\end{equation}

\noindent where 

\begin{equation}
{\bf U}={\bf u(x')-u(x)}\equiv \Delta {\bf u}
\end{equation}

\noindent 
The moments of the two-point velocity differences which
in 
homogeneous and isotropic turbulence can depend only on 
 the absolute values of two vectors
(velocity difference ${\bf v(x')-v(x)}$ and displacement 
${\bf r\equiv x'-x}$) and the angle $\theta$ between them with $\theta=\pi/2$
and $\theta=0$ corresponding to transverse and longitudinal structure 
functions, respectively. 
\noindent 
It is easy to show [2]- [3] that the 
 general form of the second-order
structure function in the inertial range is:
\begin{equation}
S_{2}(r,\theta)= \frac{2+\xi_{2}}{2}D_{LL}(r)(1-\frac{\xi_{2}}{2+\xi_{2}}cos^{2}(\theta))
\end{equation}
\noindent with $D_{LL}(r)=<(u(x)-u(x+r))^{2}>$.
 More involved relation can
 be written for the fourth-order moment:
\begin{equation}
S_{4}(r,\theta)=D_{LLLL}(r)cos^{4}(\theta)-3D_{LLNN}(r)sin^{2}(2\theta)+
D_{NNNN}(r)sin^{2}(\theta)
\end{equation}
\noindent where $D_{LLNN}=<(v(x)-v(x+r))^{2}(u(x)-u(x+r))^{2}>$
and $v$ and $u$ are the components of the velocity field perpendicular
and parallel to the $x$-axis, respectively. In general, 
in the llimit $cos(\theta)\equiv s\rightarrow \pm 1$, corresponding to the moments of the 
longitudinal velocity differences
$S_{n}(r,s)\rightarrow S_{n}(r)cos^{n}(\theta)$. 
This means that in this limit 
$Z(\lambda,r,s)\rightarrow Z(\lambda s,r)\equiv Z(\lambda_{x},r)$.
The generating function can depend only on three variables:

$$\eta_{1}=r;~~ \eta_{2}=\frac{{\bf \lambda\cdot r}}{{\bf r}}\equiv 
\lambda cos(\theta);~~ \eta_{3}=\sqrt{\lambda^{2}-\eta_{2}^{2}};$$  
The equations (7)-(10) become after some manipulations:
\begin{equation}
Z_{t}+[\partial_{\eta_{1}}\partial_{\eta_{2}}+\frac{d-1}{r}\partial_{\eta_{2}}+
\frac{\eta_{2}}{\eta_{3}}\partial_{\eta_{1}}\partial_{\eta_{3}}+\frac{\eta_{3}}{r}\partial_{\eta_{2}}\partial_{\eta_{3}}+\frac{(2-d)\eta_{2}}{r\eta_{3}}\partial_{\eta_{3}}-\frac{\eta_{2}}{r}\partial^{2}_{\eta_{3}}]Z=I_{p}+D
\end{equation}

\noindent where 
\begin{equation}
I_{p}=
\lambda_{i}<(\partial_{2,i} p(2)-\partial_{1,i} p(1))e^{\bf \lambda\cdot U}>
\end{equation}
\noindent and
\begin{equation}
D=
\lambda_{i}\nu<(\partial^{2}_{2,\alpha} v_{i}(2)-\partial^{2}_{1,\alpha}
v_{i}(1)e^{\bf {\lambda\cdot U}}>
\end{equation}

\noindent where, to simplify notation we set $\partial_{i,\alpha}\equiv
\frac{\partial}{\partial x.\alpha}$ and $v(i)\equiv v({\bf x_{i}})$.
Interested in the limit $\eta_{3}\rightarrow 0$, 
we will seek a  solution in a form:

\begin{equation}
Z=Z_{o}(\eta_{1},\eta_{2})K((\frac{\eta_{3}}{\eta_{2}})^{2}))
\end{equation}

\noindent The equation for the generating function becomes
(the subscript $o$ is omitted hereafter):

\begin{equation}
[\partial_{\eta_{1}}\partial_{\eta_{2}}+\frac{d-1}{r}\partial_{\eta_{2}}+
\frac{2\alpha}{\eta_{2}}\partial_{\eta_{1}}+\frac{2\alpha}{\eta_{2}}\frac{1-d}{r}]Z=I_{p}+D
\end{equation}

\noindent where $\alpha=K'(0)$. When $\eta_{3}\rightarrow 0$, one can introduce a more general 
relation $Z\approx 
Z_{o}(\eta_{1},\eta_{2})+\eta_{3}^{2}Z_{1}(\eta_{1},\eta_{2})$ and
 derive an equation for $Z_{1}$ [7]. This, however, does not 
 give any advantage over the expression (18).
The turbulence production contributions, which in the case of the 
large-scale forcing function rapidly varying in time 
can be written as:

\begin{equation}
I_{f}=P\eta_{2}^{2}\eta_{1}^{2}(1+2s^{2})\approx 
3\eta_{2}^{2}P\eta_{1}^{2}Z
\end{equation}
\noindent will be discussed  later. 
The form of $Z$, introduced above,
 is only an approximation,  valid in 
the $d\rightarrow\infty$ limit. One can rewrite the dissipation term 

\begin{equation}
D=-\frac{3\eta_{2}^{2}}{d}<({\cal E}(1)+
{\cal E}(2))e^{\eta_{2}U}>+s=\\
-\frac{6\eta_{2}^{2}{\cal E}}{d}Z-\frac{3}{d}\eta_{2}^{2}
<[\delta(1)+\delta(2)]e^{\eta_{2}U}>+s
\end{equation}

\noindent where the dissipation fluctuations are defined as 
$\delta(x)={\cal E}(x)-{\cal E}$. The ``surface contribution'' $s$:

$$s=\lambda_{i}\nu <\partial_{2,\mu}(\partial_{2,\mu}u_{i}(2)e^{{\bf \lambda\cdot  U}})-\partial_{1,\mu}(\partial_{1,\mu}u_{i}(1)e^{{\bf \lambda\cdot  U}})>\rightarrow 0$$

\noindent in the large Reynolds number limit $\nu\rightarrow 0$. 
Substituting this into (19) 
gives after Fourier transformation:

\begin{equation}
-\partial_{U}U\partial_{r}P-\frac{(d-1)}{r}\partial_{U}UP
+2\alpha\partial_{r} P+
2\alpha\frac{1-d}{r}P=\frac{6{\cal E}}{d}\partial^{3}_{U}P(U)+
\frac{3}{d}\partial^{3}_{U}\kappa(U)P(U)+\partial_{U}^{2}i_{p}(U)P(U)
\end{equation}

\noindent where 
\begin{equation}
\kappa(U)=<(\delta_{1}+\delta_{2})|U>
\end{equation}

\noindent and 
\begin{equation}
i_{p}=<\partial_{2,x} p(2)-\partial_{1,x}p(1)|U>
\end{equation}

\noindent are  conditional expectation values of $\delta_{1}+\delta_{2}$
and $\partial_{2,x}p(2)-\partial_{1,x}p(1)$ 
for a fixed value of the longitudinal 
velocity difference $U\equiv U_{x}$, respectively.
It is clear that

\begin{equation}
\int_{-\infty}^{\infty}\kappa(U)P(U)dU=
\int_{-\infty}^{\infty} i_{p}(U)P(U)dU=
<\delta_{1}+\delta_{2}>=<i_{p}>=0
\end{equation}

\noindent 
Multiplying (22) by $U^{3}$ gives:

\begin{equation}
\frac{1}{r^{(d-1)a}}\partial_{r}r^{(d-1)a}S_{3}=-\frac{12 {\cal E}}{d}+i
\end{equation}

\noindent where

\begin{equation}
i=6\int_{-\infty}^{\infty}i_{p}(U)UP(U)dU
\end{equation}

\begin{equation}
a=\frac{1-2\alpha/3}{1+2\alpha/3}
\end{equation}

\noindent This result 
differs from exact  Kolmogorov relation (3),  involving the factors
$r^{d+1}$ instead of $r^{d-1}$ derived above. The error stems from
the shape of the generating function (18), introduced above. In reality 
the transverse and longitudinal correlation functions are related by 
the incompressibility constraint

\begin{equation}
\frac{r}{d-1}\partial_{r}S_{2}+S_{2}=S_{2}^{t}
\end{equation}

\noindent providing a boundary condition 
 on a mixed derivatives of the $Z$-function
at $\lambda=0$. This relation is not accounted for in the anzatz (18).
 One can see that when $d\rightarrow \infty$,  the role  of the 
incompressibility
constraint and,  as a consequence,  of the pressure gradients,  
decreases leading to the 
longitudinal and transverse correlations functions very close to each other.
This is easily understood since a single constraint on the 
$d\rightarrow \infty$ velocity components cannot quantitatively 
change too much. 
It will become clear below that although, in general, 
the effect  of the pressure contributions 
is $O(1/d)$,  they cannot be 
neglected since they are responsible for the regularization of the solution, 
thus producing a  major difference between the Navier-Stokes and Burgers 
dynamics.  Comparing (26) with (3) we conclude that when $d$ is large
$a=1$, $i=0$ and that the the solution in the form 
(18) is a good approximation.

\noindent In general we know nothing about the 
function $\kappa(U)$  which is a  part of  the dissipation term $D$. 
Luckily, it is   irrelevant in the above calculation of $S_{3}$.
 To obtain a 
closed equation for the generating function,
the expression for $D$ is needed. It has been shown by
 Polyakov [4]

\begin{equation}
D=\lambda_{\mu}\nu<[\nabla_{2}^{2}u_{\mu}(2)-\nabla_{1}^{2}u_{\mu}(1)]
e^{\bf \lambda_{\beta}\cdot \Delta u_{\beta}}>
=-\frac{A}{\eta_{2}}\partial_{\eta_{1}}Z
\end{equation}
\noindent where we choose $A=\gamma d+c$ and $\gamma$ and $c$ are $O(1)$. 
The relation  (30) looks the same as the one 
appearing in the Polyakov theory of Burgers turbulence. 
This is not accidental since,  barring the pressure conntributions, the 
equation of motion (1) is formally similar to the 
multidimensional Burgers equation. The 
 expression (30 )
is a result of
an operator product expansion,  based on a point-splitting 
calculation,  developed for the one-dimensional problem of Burgers 
turbulence [3]. It is easy to show  that (30) (see Refs. [4],[5]) exactly 
satisfies the equation of
motion (1) (without pressure)  and is the only possible model,  
not violating scaling (see below) and both
Galilean invariance and an obvious constraint  $\overline U=0$. 
Substituting this into (22 ) with $\alpha=0$,  we have after 
 Fourier transformation:

\begin{equation}
\partial_{U}U\partial_{r}P+\frac{(d-1)}{r}\partial_{U}UP
=-A\partial_{r}P-\partial_{U}^{2}i_{p}(U)P(U)
\end{equation}

\noindent Keeping only the 
$O(d)$- contributions gives:

\begin{equation}
-\frac{d}{r}\partial_{U} UP=\gamma d \partial_{r}P
\end{equation}

\noindent Multiplying this by $U^{n}$ and integrating gives:
\begin{equation}
n\frac{d}{r}S_{n}=\gamma d \partial_{r}S_{n}
\end{equation}

\noindent Seeking the solution in a scaling form: 
$S_{n}\propto r^{\xi_{n}}$ 
gives 

\begin{equation}
\xi_{n}\rightarrow n/3
\end{equation}

\noindent where $\gamma=3$ is chosen to satisfy the relation (1). Thus,
the predictions of  Kolmogorov theory are valid
in the limit $d\rightarrow \infty$. One can see that in this case 

\begin{equation}
P(U,r)=\frac{1}{r^{\frac{1}{3}}}F(\frac{U}{r^{\frac{1}{3}}})
\end{equation}

\noindent which is consistent with the scaling 
assumption involved in derivation 
of the expression for the dissipation anomaly (30).
This result needs some explanation. The $d\rightarrow \infty$ limit 
must be understood in a following way: first we fix the moment number
 $n$ and then drive 
$d$ to infinity. It is clear from (31) that neglecting the pressure term
the exponents 

\begin{equation}
\xi_{n}=\frac{(d-1)n}{A-n}
\end{equation}

\noindent are singular at $n=A$ and  
violate all possible dynamic constraints. This happens due to 
the sign of the $O(1)$  contribution to the 
equation of motion (22). Thus, 
the role of the pressure
terms  is in modifying the sign of the ``Burgers- like''  first contribution 
to the left side of (22), 
thus producing the regularization,  distinguishing
Burgers dynamics from that of the Navier-Stokes. The possibilities
of modelling the pressure contributions will be discussed  
in a  future communication. It is interesting,  that gradual disappearence 
of the anomalus scaling was observed in a  numerical solution of a set
of $N$ coupled one-dimensional ``shell models'':  the deviations from 
Kolmogorov scaling decreased to close to zero with increase of $N$ [8].

Now, we can show that  the pressure
 terms $I_{p}=O(1)$ and are small compared to the $O(d)$ contributions. 
The generating function in the limit $d\rightarrow \infty$ can be written as:

\begin{equation}
Z\approx \sum (-1)^{n}A_{2n}\frac{(kr^{\frac{1}{3}})^{2n}}{n!}
\end{equation}
\noindent Assuming that at the integral scale $r=L=1$ 
the PDF becomes close to the gaussian gives $A_{2n}=2^{n}(2n-1)!!$:

$$Z=<e^{\lambda U}>\approx e^{2\lambda^{2}r^{\frac{2}{3}}}$$
\noindent and 

$$\sqrt{<e^{2\lambda U}>}\leq Z$$

\noindent since $\lambda=ik$ with real $k$.  Using this relation the 
Schwartz inequality is:
\begin{equation}
I_{p}\leq \sqrt{\overline{({\bf \lambda\cdot \nabla}\frac{\nabla_{\alpha}
\nabla_{\beta}}{\nabla^{2}}v_{\alpha}v_{\beta})^{2}}}~Z
\end{equation}
\noindent The four- order correlation function 
$\overline{v_{\alpha}v_{\beta}v_{\gamma}v_{\delta}}
\approx 
<v_{\alpha}v_{\gamma}><v_{\beta}v_{\delta}>+
....$ where

$$\overline{v_{\alpha}v_{\gamma}}\approx v^{2}_{rms}(r^{2}
\delta_{\alpha,\gamma}-\frac{2r_{\alpha}r_{\gamma}}{d+1})$$

\noindent with $v^{2}_{rms}=O(1)$. It means that 
$\nabla_{\alpha}\overline{v_{\alpha}v_{\beta}}= 
\nabla_{\beta}\overline{v_{\alpha}v_{\beta}}= 0$. 
The only sorce of the $O(d)$ contributions is $\nabla_{i}r_{i}=d$ which
cannot appear in (38) due to incompressibility. That is why 
$I_{p} \leq 
 QZ$,  where $Q$ is the
operator,  involving $\eta_{1}$, $\eta_{2}$ and derivatives, all $O(1)$.

\noindent 
To conclude: Kolmogorov scaling is an exact solution of the Navier-
Stokes equations in the limit of large space dimensionality
$d\rightarrow \infty$,  provided the assumptions leading to
a self- consistent conjecture for the 
dissipation anomaly, similar to the one derived by Polyakov [4] in the
theory of Burgers turbulence, are valid. We do not know if this solution 
is the only possible one. 
The $O(1/d)$ pressure contributions to the 
Navier-Stokes equations, though unable to
strongly influence the values of the scaling exponents, are crucial 
providing the regularization of the otherwise singular expression for the 
scaling exponents. These terms are responsible for 
a clear distinction between 
the Navier-Stokes and Burgers dynamics. 

\noindent The equation (7) can be written for the multidimensional Burgers 
equation if density fluctuations are taken into account [4],[5]. The fact 
that in the limit $d\rightarrow \infty$,
 the pressure terms are 
unimportant make the Navier-Stokes and Burgers equations formaly seem
the same. This not so. The steady state solution 
in the random- force-driven Burgers 
equation is
impossible without accounting  for  the forcing term (20) which 
imposes  a typical Burgers scaling 
behaviour $U=O(r)$ [4],[5]. This scaling is not always there: 
It has been shown in [9] and [10] that,  in principle,
the scaling of velocity correlation functions in Burgers 
turbulence depends on the properties of the forcing function. 
In the case of the Navier-Stokes dynamics 
one can drop  the forcing term and obtain the inertial range Kolmogorov
scaling $U=O(r^{\frac{1}{3}})$. However, as was shown above, neglecting the
pressure contributions leads to the singularity of some high-order moments
and to the unphysical behaviour. Still,  when 
$d$ is large, due to  formal similarity
of the equations of motion, one cannot rule out  Kolmogorov scaling 
as a short-time asymptotics of  decaying Burgers turbulence.

\noindent I am grateful to S. Boldyrev, M. Chertkov,
R.H.Kraichnan, A. Polyakov, B. Shraiman and M. Vergassola for their help 
and most stimulating and interesting discussions.

\noindent {\bf references}
\\
1. U. Frisch, J.-D. Fournier and H. Rose, J.Phys.A, {\bf 11} 187 (1978)
\\
2.  L.D.Landau and E.M. Lifshitz, Fluid Mechanics, Pergamon Press, Oxford, 1987;\\
3. A.S.Monin and A.M.Yaglom, ``Statistical Fluid Mechanics'' vol. 1, MIT Press,
Cambridge, MA (1971)
\\
4.  A.M. Polyakov, Phys.Rev. E {\bf 52}, 6183 (1995)
\\
5. S. Boldyrev, Phys.Rev.E {\bf 55}, 6907 (1997)
\\
6. V. Yakhot, Phys.Rev. E, (1998)
\\
7. I am grateful to S. Boldyrev for this suggestion.
\\
8. L. Biferale and D. Piorelli, private communication (1998)
\\
9. A. Chekhlov and V. Yakhot, Phys.Rev.E {\bf 51}, R2739 (1995);
\\
10.  V. Yakhot and A.Chekhlov, Phys.Rev.Lett. {\bf 77}, 3118 (1996)
\\

\end{document}